\documentclass[article,11pt,onecolumn,showpacs,preprintnumbers,amsmath,assume,plain]{revtex4}
\usepackage{graphicx}
\usepackage{bm}
\providecommand{\abs}[1]{\vert#1\vert}
\usepackage{float}
\usepackage{hyperref}

\begin{document}
\begin{titlepage}
\title{Quasi-point versus point nodes in $Sr_2RuO_4$, the case of a flat tight binding $\gamma$ sheet}.

\author{P. Contreras$^{1}$, Dianela Osorio$^{1,2}$, and Shunji Tsuchiya$^{3}$}
\affiliation{$^{1}$ Department of Physics, Universidad of Los Andes, Merida, Venezuela}
\affiliation{$^{2}$ Actually at Dept of Brain and Behavioral Sciences, University of Pavia, Pavia, Italy}
\affiliation{$^{3}$ Department of Physics, Chuo University, Tokyo, Japan}

\date{\today}

\begin{abstract}
We perform a numerical study of the unitary regime as a function of disorder concentration in the imaginary part of the elastic scattering cross-section for the compound $Sr_2RuO_4$ in the flat band non-disperse limit. By using a self-consistent tight binding (TB) method, we find a couple of families of Wigner probabilistic functions that help to explain macroscopically the distribution between Fermion dressed quasiparticles and Cooper pairs, and also the position of nodes in the order parameter for $Sr_2RuO_4$. Therefore, we are able to show that a TB model for the FS $\gamma$-sheet, numerically shows 4 point nodes in a flat $\gamma$ sheet limit, or 4 quasi-point nodes for strong dispersion $\gamma$ sheet limit in the reduced phase scattering space (RPS).

{\bf Keywords:} Unconventional superconductivity; Flat bands, Triplet reversal time broken state; FS $\gamma$-sheet; $Sr_2RuO_4$; non-magnetic disorder; Elastic scattering cross-section; Tiny gap, Point nodes, Quasi-point nodes, Wigner distribution probabilities, Computational material design. 
\pacs{{34.80.Bm; 72.10.-d; 72.80.Ng; 74.20.-z; 74.62.Dh; 74.70.Dd; 74.70.-b} \vspace{-5pt}}
\end{abstract}

\maketitle
\end{titlepage}

\section{Introduction}

Strontium ruthenate ($Sr_2RuO_4$) [1], a ternary body-centered tetragonal crystal with a layered square structure for the ruthenium atoms has a normal state described by a Fermi liquid model [2], with three FS metallic conduction sheets, namely the $\alpha$ and $\beta$ (1D), and $\gamma$ (2D) ones. Moreover, $Sr_2RuO_4$ belongs to the family of low temperatures unconventional superconductors, with $T_c\approx$ 1.5 K, but strongly depending on non-magnetic disorder [3]. Furthermore, $Sr_2RuO_4$ is an unconventional superconductor with triplet pairing and some type of nodes in the order parameter for each sheet of the FS [4]. In addition, the symmetry of the superconducting gap breaks the time reversal symmetry [5-7].

The discovery of several superfluid phases at ultralow temperatures in the liquid isotope $^3 He$ in 1972 by Lee, Oshero and Richardson was an initial point of remarkable investigations in unconventional superconductors [8]. $^3 He$ atoms are naturally fermionic, and thus the formation of Cooper pairs in $^3 He$, provides a system similar to bulk superconductors but with some differences. $^3 He$ is superconducting in a liquid state at ultralow temperatures with $T_c \sim 2.6$ mK, the order parameter (OP) shows an odd momentum k-pairing dependence in one of the phases of $^3 He$, namely in the A phase with a p-wave spin triplet broken state with jumps observed in specific heat in an external magnetic field [9]. 

In 1994 the discovery of the low temperature unconventional bulk superconductor $Sr_2RuO_4$ has since then attracted a lot of attention [1] due to the similarities with the A phase of the isotope $^3 He$ [4]. Strontium (stoichiometry) ruthenate is isostructurally similar to the the HTSC compounds but without copper. In contrast to the HTSC, it shows triplet odd pairing with a p-wave order parameter and k-dependence similar to the OP proposed in phase A of fermionic $^3 He$. A recent review favoring this argumentation is given in [38]. 

We use the elastic scattering cross-section in our work to study strontium ruthenate in order to obtain a classical window [18] to the quantum effects in the unitary limit [41] for the FS $\gamma$-sheet, therefore we consider our work important, since this approach was previously not considered to be relevant in this compound, even it was used widely for HTSC in the past, mainly by Carbotte and collaborators [22]. Additionally, we use a couple of TB Wigner distribution probabilities aiming at helping to clarify and to contrast visually the location of the nodes.

This communication concerts the 2D $\gamma$ sheet of the FS in $Sr_2RuO_4$. We investigate using a tight binding approach (TB) [10], the conjecture of nodes position, due to the fact that it continues to be a matter of intense discussions among the scientific community, despite it was experimentally discovered 27 years ago [11-17] and references there in. 

As in previous works using a Wigner probabilistic distributions macroscopic approach [18], in this numerical study, we use a TB first nearest neighbor expression for the dispersion law $\xi_\gamma(k_x,k_y)$ in order to model the FS $\gamma$ sheet, which is centered at (0,0) in the first Brillouin zone. We extend our previous phenomenological works [119-20] by varying and analysing the behavior of one of the TB parameters, i.e., specifically the Fermi energy $|\epsilon_F|$ accordingly to table 1.

Following this idea, we have noticed that by making the Fermi energy TB parameter $|\epsilon_F|$ very small in absolute value for the $\gamma$ sheet, we get point nodes (shown in Figure 2) in the MN model instead of quasi-point nodes of Figure 1. Therefore, the case shown in Figure 2 corresponds to the half-filling metallic limit with one electron per site, and with almost not dispersion, the so called flat band limit, and where the ground state of $Sr_2RuO_4$ has the lowest $N$ states occupied. In this case, superconducting $Sr_2RuO_4$ is therefore gapless, i.e., is is a metallic unconventional system, meaning that it can be excited above the ground state by any infinitesimal energy/temperature value, and it correspond to a point nodes model as we numerically aim at showing in this work.

We also notice, that the quantum mechanics dispersion effect is caused by the hopping integral between local orbitals, is the one that makes the energy of the $\gamma$ sheet with values $|t| \approx |\epsilon_F|$ to have quasi-node points, but as it was note in [21] in some cases, i.e., with only first neighbors hopping there are bands with no dispersion, and these bands are called flat bands. It is very interesting to notice that in reference [21], it is state that these compounds do not include magnetic elements, such as transition or rare-earth elements, but we point out that the element Strontium ($Sr$) is non-magnetic, therefore is not excluded from their approach. 

\begin{table}
\begin{tabular}{|c c c c|} 
 \hline
 ( t, \; $\epsilon_F$, \; $\Delta_0^{\gamma}$ ) \hspace{0.6cm}  & type of nodes \hspace{0.6cm} &  dispersion law limit \hspace{0.6cm} & Figure \\ [0.5ex] 
 \hline\hline
 (0.4, 0.4, 1.0) (meV) & 4 quasi-point nodes & strong disperse law & Figures 1 and 3 ($|t| \approx |\epsilon_F|$) \\ 
 \hline
 (0.4, 0.04, 1.0) (meV) & 4 point nodes & flat non-disperse law & Figures 2 and 4 ($|t| \gg |\epsilon_F|$) \\ [1ex] \hline
\end{tabular}
\caption{TB Parameter and dispersion law values and limits for the quasi-point and point models nodes in the $\gamma$ sheet of $Sr_2RuO_4$}
\label{table:1}
\end{table}

In this research, we set up a new numerical study with almost non dispersion in $\xi_\gamma(k_x,k_y)$, i.e., when $|t| \gg |\epsilon_F|$, in contrast with our previously studied [19-20] where we used $|t| \approx |\epsilon_F|$.

In other order of ideas, the 2D TB normal state electronic energy expression in a first neighbor approximation is given by $\xi_\gamma(k_x,k_y)= -\epsilon_F + 2t \; [cos(k_x a)+cos(k_y a)]$,  which follows electron-hole symmetry. The 2D TB OP expression in a first neighbor approximation, corresponds to the MN model [16]. These equations allow the study of the triplet time symmetry broken state in the FS $\gamma$-sheet of $Sr_2RuO_4$, i.e., $\bf{\Delta^\gamma}$ $(k_x,k_y)$ $=$ $\Delta_0 \; \bf{d}^\gamma(k_x,k_y)$, with $\bf{d^\gamma}$ $(k_x,k_y)$ $ = [(sin(k_x a) + i \; sin(k_ya)]\,\bf{z}$ and $\Delta_0^{\gamma}$ = 1.0 meV, according to experimental measurements in impurity samples [3].

\begin{figure}[ht]
\includegraphics[width = 5.0 in, height= 5.0 in]{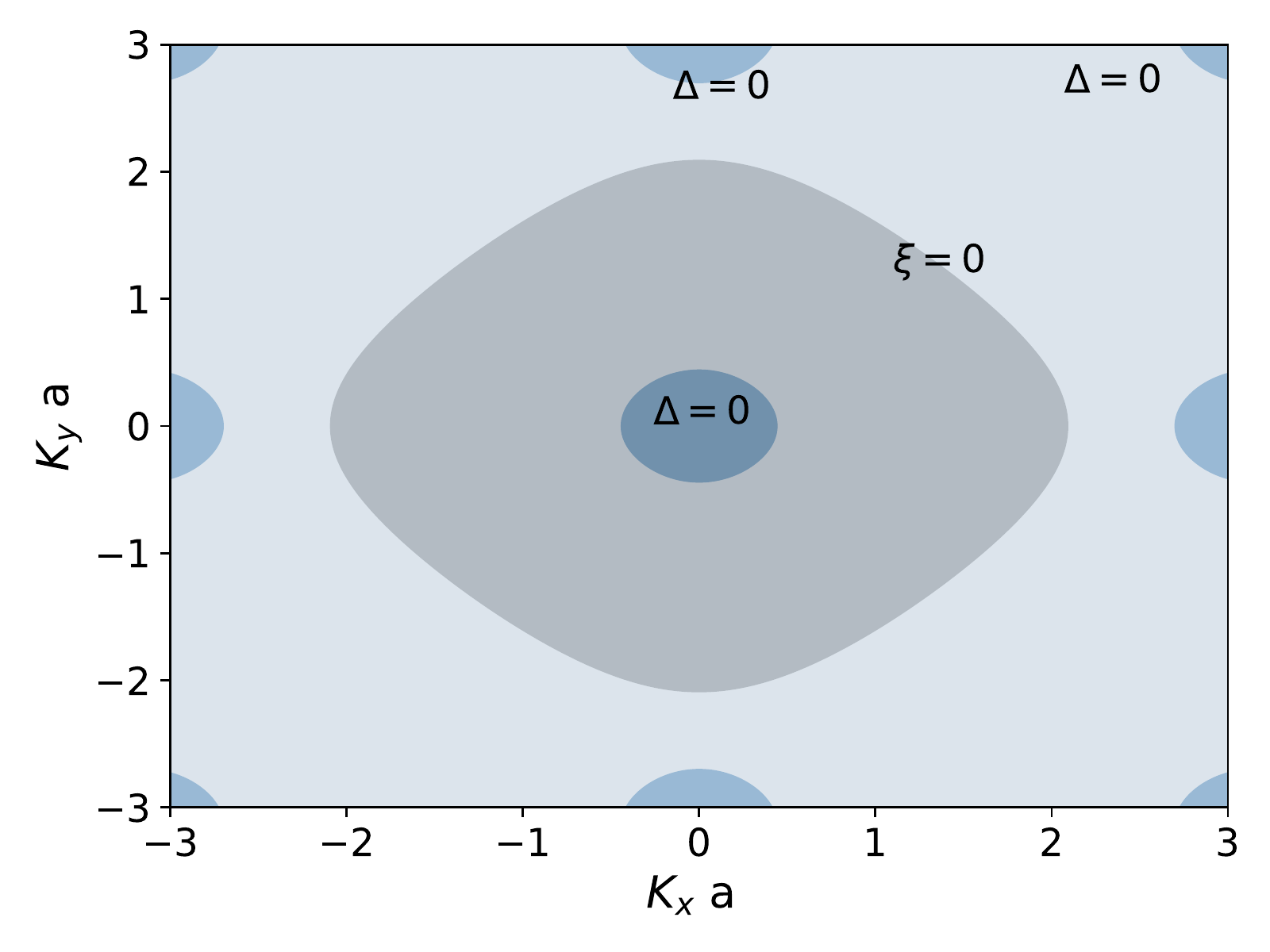}
\caption{2D implicit plot of the TB anisotropic Fermi $\gamma$ sheet $\xi_\gamma(k_x,k_y)$ = 0 and the triplet superconducting Miyake-Narikiyo tiny gap with the localization of the nine quasi-point nodes where $\bf{d}^\gamma (k_x, k_y)$ = $\bf{0}$ from the TB values with $|\epsilon_F|$ $\approx$ $|t|$. This model gives an electronic anisotropic dispersion law for the normal state quasiparticles.}
\end{figure}

The nine points where the order parameter $\bf{d}^\gamma(k_x,k_y)$ has zeros are sketched in both, figure 1 and figure 2. From nine OP zeros only 4 points, symmetrically distributed in the \{10\} and \{01\} planes at $k$-points $(0,\pm\pi)$ and $(\pm\pi,0)$ give symmetric point nodes, according to group theoretical considerations, when $\epsilon_F << 1$ and when $|t| \gg |\epsilon_F|$. The other five points in figure 1, do not touch the $\gamma$ sheet, i.e., the 4 points symmetrically distributed in the \{11\} planes at $k$-points $(\pm\pi,\pm\pi)$ and 1 point in the \{00\} plane at $k$-point (0,0). 

As noticed firstly in [16] and also by us in [19-20], the gap on the $\gamma$ sheet is very anisotropic and leaves a tiny gap $\Delta_0^{\gamma}$ around four points, $(0,\pm\pi)$  and $(\pm\pi,0)$, but now we state that when $|t| \gg |\epsilon_F|$ there is not such a gap. According to group theory considerations, in this case as in the tiny gap MN model, the imaginary OP has two components which belong to the irreducible representation $E_{2u}$ of the tetragonal point group $D_{4h}$ [17]. It also corresponds to a triplet odd paired state $\bf{d}^\gamma (-k_x,-k_y)= - \bf{d}^\gamma (k_x,k_y)$ with the same basis functions $sin(k_{x} a)$ and $sin(k_{y} a)$ and Ginzburg-Landau coefficients $(1,i)$ [17,39]. 

\begin{figure}[ht]
\includegraphics[width = 5.0 in, height= 5.0 in]{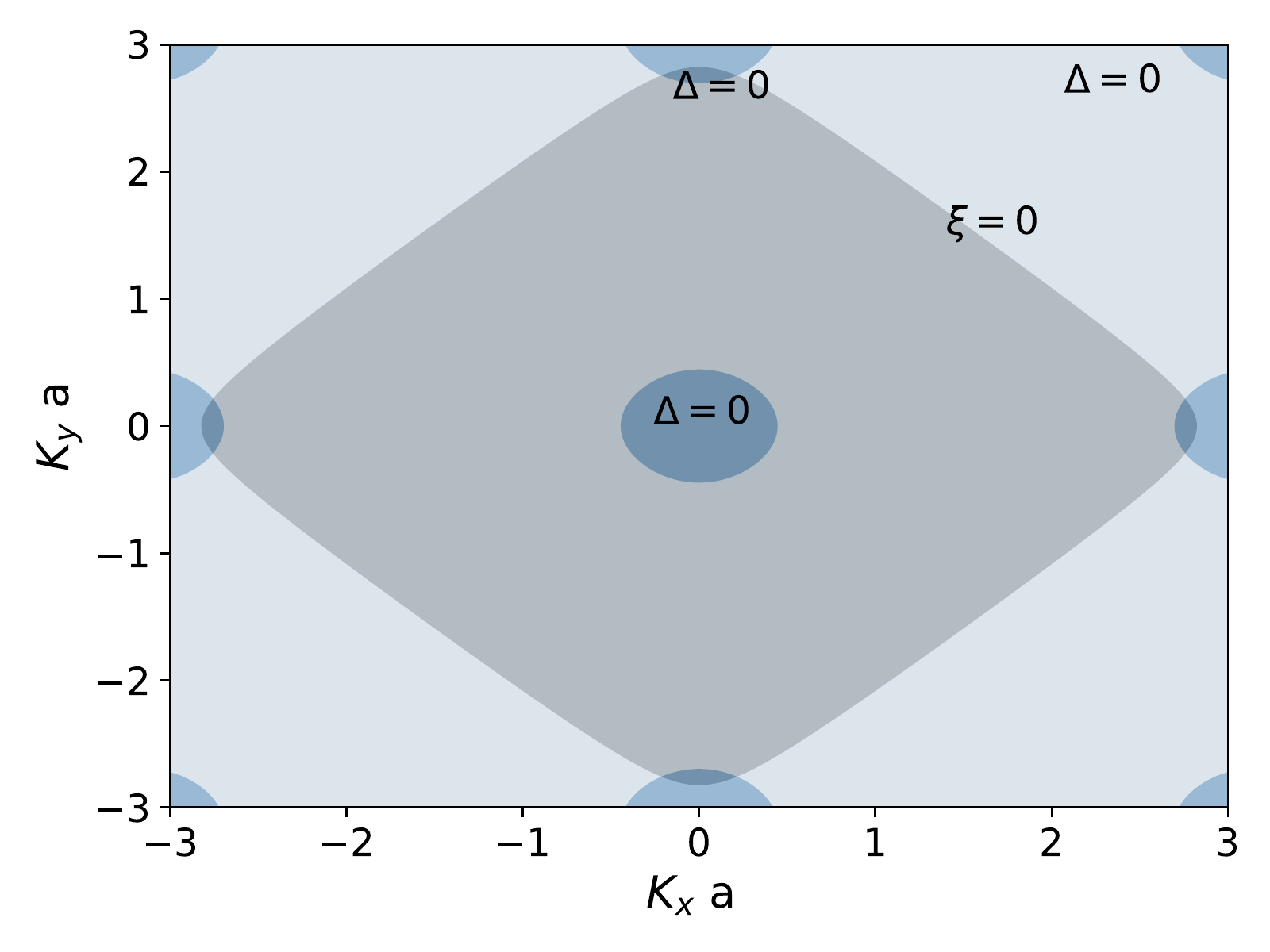}
\caption{2D implicit plot of the tight binding anisotropic Fermi $\gamma$ sheet $\xi_\gamma(k_x,k_y)$ = 0 and the triplet superconducting Miyake-Narikiyo point nodes half filling state with the localization of the nine points where $\bf{d}^\gamma (k_x, k_y)$ $=$ $\bf{0}$ from the TB values with $|\epsilon_F|$ $\approx$ $\frac{|t|}{10}$. This model gives a flat electronic dispersion law for the normal state quasiparticles.}
\end{figure}

We ought to emphasize that the difference with our previous work [19], consists in using the metallic ground state, which is given by decreasing an order of magnitude of the Fermi energy for the $\gamma$ sheet, with a new value $\epsilon_F = 0.04$ meV, as can be observed in figure 2, giving us a new model with natural point nodes, that intercept the gap in 4 points on the $\gamma$-sheet, making it gapless, namely, $(0,\pm\pi)$ and $(\pm\pi,0)$ arriving to a 2D non disperse flat $\gamma$ sheet model for $Sr_2RuO_4$ .

We use a numerical methodology able to control the non-magnetic disorder values, proposed by J. Carbotte and E. Schachinger [22], i.e., we vary first the inverse of the strength parameter $c$ from 0 to 1 and second, we vary the value of the parameter concentration $\Gamma^+$, from optimal to dilute doping in the function in the imaginary part of the scattering cross-section. The numerical analysis is performed in the reduced elastic scattering phase-space (RPS), giving us two families of Wigner distribution macroscopic probabilities. This methodology resembles an analysis using quantum collision theory with a phase-space scattering space with a Wigner probabilistic distribution, where those represent a classical window to study the quantum world [18]. This numerical approach agrees with a previous theoretical formalism, such as the work developed by I. M. Lifshitz and collaborators on disordered systems [23].

In the following sections, we report first, a theoretical derivation of the scattering cross-section formalism, and second, a visual numerical analysis of the imaginary part of the elastic scattering cross-section on $Sr_2RuO_4$ at the phenomenological level, using the gapless point nodes model of figure 2 and comparing it, with the quasinodal case previously studied [19-20]. 

\section{Non-magnetic impurity scattering, the Edwards-Nambu-Gorkov formalism}

It is known that non-magnetic elastic scattering destroys the coherence of the Cooper pairs in unconventional superconductors. The decrease of $T_c$ is a function of the concentration of non-magnetic impurities ($n_ {imp}$). Thus, the maximum transition temperature $T_c$ is given for the case when $n_{imp} = 0$ and is denoted as $T_c^0$.

S. Edwards in 1961 (see [24] for a summary and original references) introduced the technique that allowed the use of Feynman diagrams to include the Hamiltonian that would describe the effect of static impurities in a normal metal, his model assumes the following physical conditions: at $T = 0$, the $N$ impurities are equal and independent, they are randomly distributed in the normal metal (however on a macroscopic scale the metal is homogeneous), and finally, a very important point, this is a quantum model, since impurities scatter electrons elastically and there is no loss of energy in collisions.

A dressed normal metal Green function ($GF$) is defined as an averaged by non-magnetic impurities $GF$, in Fourier components and its self-consistent form is given as follows

\begin{equation}
\displaystyle\left\langle G(\mathbf{k}, \varepsilon) \displaystyle\right\rangle = G^0(\mathbf{k}, \varepsilon) + G^0(\mathbf{k}, \varepsilon) \bm{\Sigma}(\mathbf{k}, \varepsilon)\displaystyle\left\langle G(\mathbf{k}, \varepsilon) \displaystyle\right\rangle.
\label{GreenAC}
\end{equation}

In this case, the average represents a sum over all the electronic momenta $\mathbf{k}$ and for all the points $j$ locating at each of the $N$ impurities.

\begin{equation}
\left\langle...\right\rangle = \displaystyle\sum_{k,j} 
\end{equation}

The solution to Eq. (\Ref{GreenAC}), which represents the $GF$ normalized by the effects of the scattering by non-magnetic impurities is

\begin{equation}
\displaystyle\left\langle G(\mathbf{k}, \varepsilon) \displaystyle\right\rangle = \displaystyle\left(\frac{1}{G^0(\mathbf{k}, \varepsilon)} - \bm{\Sigma}(\mathbf{k}, \varepsilon) \displaystyle\right)^{-1},
\label{promediadoG}
\end{equation} 
with an undressed normal metal $GF$ given by: 
\begin{equation}
G^0(\mathbf{k}, \varepsilon) = \displaystyle\frac{1}{i\varepsilon - \xi_{\mathbf{k}}}.
\label{green}
\end{equation}

On the other hand, the $GF$ in a superconductor for a Cooper pair in the presence of a  non-magnetic impurity potential is known as the Edwards-Nambu-Gorkov $GF$ [25] and is defined as 

\begin{equation}
\hat{G}(\varepsilon,\textbf{k}) = \hat{G}_0(\varepsilon,\textbf{k}) + \hat{G}_0(\varepsilon,\textbf{k})\hat{T}(\varepsilon)\hat{G}_0(\varepsilon,\textbf{k}). 
\end{equation}

The symbol $\wedge$ introduced by Y. Nambu above the $GF$ means that the superconducting state is described by a two-dimensional Pauli Matrix basis, since it takes into account the spin space, in addition to the momentum space. Any function in the spin space can be decomposed in terms of the Pauli matrices, and it becomes a scalar for a normal metal [24]. 

The normal $GF$ in the superconducting state is given by the expression $\hat{G}_0(\varepsilon,\mathbf{k})$ (We will not use in this work, the anomalous $\hat{F}(\varepsilon,\mathbf{k})$ Green function introduced by L. Gorkov [25] and references therein, since the anomalous $F$ Green function has the kernel $\Delta$ - gap as a function of temperature, and it is useful for a study of impurities temperature dependence on the gap, which is not the scope of this research), therefore $\hat{G}_0(\varepsilon,\mathbf{k})$ is written as

\begin{equation}
\hat{G}_0(\varepsilon,\mathbf{k}) = \frac{\varepsilon\hat{\sigma}_0 + \xi_{\mathbf{k}}\hat{\sigma}_3 + \Delta(\mathbf{k})\hat{\sigma}_1 }{\varepsilon^{2} - \xi_{\mathbf{k}}^{2} - {\abs{\Delta(\mathbf{k})}}^{2}}.
\label{normImp}
\end{equation}

The array $\hat{T}$ has the following general form

\begin{equation}
\hat{T}(\varepsilon) = \hat{U}_0\cdot\displaystyle\left[\hat{\sigma}_0 - \hat{U}_0\displaystyle\sum_\mathbf{k}\hat{G}_0(\varepsilon,\mathbf{k})\displaystyle\right]^{-1},
\label{tmatriz}
\end{equation}
with the impurity potential matrix $\hat{U}_0 = U_0 \hat{\sigma}_3 $. Since we want to use a tight binding numerical formalism, it is convenient to define an average where the sums in the space $\mathbf{k}$ will be now averages over the Fermi Surface for each component of the $GF$ is described by the Pauli matrices in the following way: 

\begin{equation}
\displaystyle\sum_\mathbf{k}\hat{G}_0(\varepsilon,\mathbf{k}) = \displaystyle\sum_\mathbf{k} \displaystyle\left( G_{0}^0(\varepsilon,\mathbf{k})\hat{\sigma}_0 + G_{0}^1(\varepsilon,\mathbf{k})\hat{\sigma}_1 + G_{0}^3(\varepsilon,\mathbf{k})\hat{\sigma}_3  \displaystyle\right).
\label{Gcomp}
\end{equation}

Studying each component separately we have the following relations:

\begin{equation}
 \displaystyle\sum_\mathbf{k} G_{0}^0(\varepsilon,\mathbf{k}) = -i\pi N_F\left\langle  \frac{\varepsilon}{\sqrt{\varepsilon^{2} - {\abs{\Delta(\mathbf{k})}}^{2}}}\right\rangle_{FS},
\label{g0}
\end{equation}

\begin{equation}
 \displaystyle\sum_\mathbf{k} G_{0}^1(\varepsilon,\mathbf{k}) = -i\pi N_F\left\langle  \frac{\Delta(\mathbf{k})}{\sqrt{\varepsilon^{2} - {\abs{\Delta(\mathbf{k})}}^{2}}}\right\rangle_{FS},
\label{g1}
\end{equation}
and
\begin{equation}
 \displaystyle\sum_\mathbf{k} G_{0}^3(\varepsilon,\mathbf{k}) = -i\pi N_F\left\langle  \frac{\xi_{\mathbf{k}}}{\sqrt{\varepsilon^{2} - {\abs{\Delta(\mathbf{k})}}^{2}}}\right\rangle_{FS}.
\label{g3}
\end{equation}

Integrating over the space $\mathbf{k}$ for the energy of the metal and averaging in our TB model on the Fermi surface, assuming electron-hole symmetry, we have two important conditions: $\sum_\mathbf{k} G_{0}^1(\varepsilon,\mathbf {k}) = 0$ for a superconductor in a triplet state with D$_{4h}$ symmetry, since the order parameter has odd parity $\Delta^\gamma_{-k} = - \Delta^\gamma_{k}$. In addition, the value of $\sum_\mathbf{k}G_ {0}^3(\varepsilon,\mathbf{k}) = 0$ means that in the metallic system there is a symmetric electron-hole physical system $\xi^\gamma_k = \xi^\gamma_{-k}$, as we pointed out in the introduction. With these two conditions, we rewrite (\ref{Gcomp}) with a new function that we call $g(\varepsilon)$

\begin{equation}
\displaystyle\sum_\mathbf{k}\hat{G}_0(\varepsilon,\mathbf{k}) = \displaystyle\sum_\mathbf{k} G_{0}^0(\varepsilon,\mathbf{k})\hat{\sigma}_0 = -i \; \pi \; N_F \; g(\varepsilon) \; \hat{\sigma}_0,
\label{Gcomp2}
\end{equation}
where $g(\varepsilon)$ is given by the equation

\begin{equation}
g(\varepsilon) =\left\langle\frac{\varepsilon}{\sqrt{\varepsilon^{2} - {\abs{\Delta(\mathbf{k})}}^{2}}} \right\rangle_{FS}.
\end{equation}

Now with this result, we calculate the inverse matrix of the array $\hat{T}$ 

\begin{equation}
\hat{T}(\varepsilon) = U_0\hat{\sigma}_3\cdot\displaystyle\left[\hat{\sigma}_0 + i\pi N_FU_0\hat{\sigma}_3\cdot(g(\varepsilon)\hat{\sigma}_0)\displaystyle\right]^{-1},
\label{tmatriz1}
\end{equation}
and,
\begin{equation}
\displaystyle\left[\hat{\sigma}_0 + i\pi N_FU_0g(\varepsilon)\hat{\sigma}_3\displaystyle\right]^{-1} = \displaystyle\frac{\displaystyle\left[\hat{\sigma}_0 + i\pi N_FU_0g(\varepsilon)\hat{\sigma}_3\displaystyle\right]}{1 -(i\pi N_FU_0g(\varepsilon))^2 } 
\end{equation}
to get
\begin{equation}
\hat{T}(\varepsilon) =  \displaystyle\frac{U_0\hat{\sigma}_3 + i\pi N_F{U_0}^2g(\varepsilon)\hat{\sigma}_0}{1 - (i\pi N_FU_0g(\varepsilon))^2}.
\end{equation}

Instead of using the impurity potential $U_0$, we use the parameter $c$ inverse to the strength $U_0$. The parameter $c$ which is defined as $c = (\pi\,N_F\,U_0)^{-1} = \tan \delta_0$ is related to the phase shift $\delta_0$ of the Cooper pair wave function. Recalling that by assuming e-h symmetry, it is obtained that the component $T^3(\varepsilon) = 0$, and 

\begin{equation}
\hat{T}(\varepsilon) =  \displaystyle\frac{ i (\pi N_F)^{-1}g(\varepsilon)}{c^2 + \abs{g(\varepsilon)}^2 }\hat{\sigma}_0.
\end{equation}

The self-energy induced by the concentration of impurities $n_ {imp}$ is denoted by $\bm{\Sigma}(\epsilon) = n_{imp} T(\epsilon)$, where $\Gamma^+ = n_ {imp}/(\pi \; N_F)$, and where finally the renormalized self-energy due to the presence of non-magnetic impurities is given by

\begin{equation}
t(\varepsilon) = \varepsilon + \bm{\Sigma}(t(\varepsilon)).
\label{autoE1}
\end{equation}

Since we use Planck units $(\hbar = k_B = c = 1)$, then Eq.(\Ref{autoE1}) can be converted according to the following self-consistent expression which finally, allow us to have an expression for the scattering cross-section parameterized as function of $c$ and $\Gamma^+$ [22]

\begin{equation}
\tilde \omega \big(\;\omega+i0^+)\; = \omega + i\pi\Gamma^+ \displaystyle\frac{g(\tilde{\omega})}{c^2 + \abs{g(\tilde{\omega})}^2}
\label{omegaTildeImp}
\end{equation}

The first thing to notice is that they represent a classical set of Wigner distribution probabilities, the second issue about (\ref{omegaTildeImp}) is that it describes both the Born classical elastic scattering and also the coherent phase (unitary) limit for which $c = 0$. Thirdly, it should be noted that the real part of the equation represents the energy $\hbar\omega$. Fourthly, the imaginary part of eq. \ref{omegaTildeImp} defines the inverse of the quasiparticle dressed lifetime $\tau^{-1}(\omega)$ in a RPS according to the expression

\begin{equation}
   \label{eq:2}
   \tau^{-1} \; (\omega) \;  = 2 \; \Im \; [ \;\tilde{\omega} \; \big( \; \omega+i0^+) \;].
\end{equation}

Therefore, we have theoretically derived the main equation for the elastic scattering cross-section formalism $\tilde{\omega}\,\big(\,\omega\,+i\,0^+)$ in the case of non-magnetic disorder, in order to model low energy self-consistent frequencies in the unitary region of the reduced scattering phase space, by varying the tight binding parameters, to obtain several families of the macroscopic Wigner probability distributions. 

For very large values of $U_0$, the unitary limit $(\ell \; k_F \sim l \; a^{-1} \sim 1 \; \& \; c = 0)$ in eq. \ref{omegaTildeImp} is given by the expression 
\begin{equation}
   \label{eq:3}
   \tilde{\omega} \big(\omega+ i 0^+) = \omega + i \pi \Gamma^+ \frac{1}{g(\tilde{\omega})}
\end{equation}

The function $g(\tilde{\omega})$ in eq. \ref{omegaTildeImp} is given by \[
        g(\tilde{\omega}) = \Bigg{\langle} \frac{\tilde{\omega}}{\sqrt{\tilde{\omega}^{2} - |\Delta|^2(k_x,k_y)}} \Bigg{\rangle}_{FS},
\]
and the average over the $\gamma$ sheet of the FS $\langle\ \ldots \rangle_{FS}$ is performed following an integration over the FS according a numerical technique successfully used to fit experimental low temperatures data with an accidental 3D point nodes TB model, i.e, the ultrasound attenuation, the electronic heat transport, and the electronic specific heat in $Sr_2RuO_4$ [26-28].
This kind of approximation is indeed characterized by root singularities at the edges of the dependence for the unitary case, i.e., where $g(\tilde{\omega})$ can go to zero within the model, following Lifshitz disorder system theory [23].

Born's approximation applies and if $c \gg$ 1 (i.e. $U_0 \ll$1) with a disorder renormalized doping parameter $\Gamma^+_B$ $= \Gamma^+/c^2 \ll 1$, with is proportional to the square of the strength potential $U_0$, and to $n_{imp}$. However, as we recently reported [20], Born scattering does not play a role in the low temperature properties of $Sr_2RuO_4$, it has a RPS window of $\Delta_0 = 1.0 \; meV$. At this point we also emphasize, that for the case of the lines nodes superconductor $La_{2-x}Sr_xCuO_4$, we reported that this compound can be numerically be in the Born limit with a RPS scattering window if $\Delta_0 = 33.9 \; meV$, for it we found another family of Wigner probabilistic distributions [20,29]. 

On the other hand, the unitary limit has a unique feature which is the resonance at zero frequency, that is, $\tilde{\omega}\,(0)= i \, \gamma$, where $\gamma$ defines the "impurity averaged" zero energy elastic scattering rate [22], and determines the crossover energy scale separating several scattering limits (Born, intermediate and unitary). Finally, impurity effects within unitary scattering has been widely investigated in several works [30-36] among other references, but not exactly within the TB macroscopic Wigner probabilistic distribution approach, we perform here.

\section{The unitary limit for the case of quasinodal points in the FS gamma-sheet of Strontium ruthenate}

In this section, we calculate numerically and visualize the behavior of the imaginary part of the elastic scattering cross-section in the unitary limit ($c = 0$) for different values of the impurities concentration parameter $\Gamma^+$, starting at very dilute disorder (turquoise line), to an optimal disorder (gray line) as it was done in our previous publications [19-20] but with a different purpose, the study of the flat $\gamma$ sheet limit. We use the following TB parameters $(|t|, |\epsilon_F|) = ( 0.4, 0.4) \; meV$ given in table 1, where there is anisotropic dispersion reflected in the normal state electronic energy that can be seen in figure 1.

The unitary limit means that the dressed normal state quasiparticles have an ill-defined momentum between elastic collisions, but the energy is conserved [41]. In addition, the signature of the unitary state is the resonance at zero frequency with the parameter $c$ = 0, in the imaginary part of the scattering cross-section as can be seen from all Wigner distribution functions of figure 3.

If $|\epsilon_F|$ $\approx$ $t$, as in figure 1, the unitary regime in the elastic scattering due to non-magnetic disorder is so strong that the mean-free path $\ell$ becomes comparable to the inverse Fermi momentum $k_F^{-1}$, and to the lattice parameter $a$, and it has two macroscopic phases, the tiny gap phase ($\Gamma^+ = 0.05$ meV - dilute levels of non magnetic disorder, turquoise line and region in phase scattering space shaded turquoise in figure 3) where the energy interval between 0.85 meV and 1.0 meV is composed only by Cooper pairs $(\tilde{\omega} \big(\omega+ i 0^+) = 0 \; meV)$, and the mixed phase with both, Cooper and normal state quasiparticles for the whole range of energies from 0 to 4 meV (from pale blue to gray color lines and a maximum region in scattering space shaded gray in figure 3 and $\Gamma \geq 0.10$ meV ). 

This corresponds to the quasinodal point nodes model as was previously reported in [19-20]. We also observe the smooth resonance centered at zero frequency for all values, with smaller values of residual zero energy $\gamma$ for very dilute values of disorder $\Gamma^+$. 

\begin{figure}[ht]
\includegraphics[width = 6.0 in, height= 5.0 in]{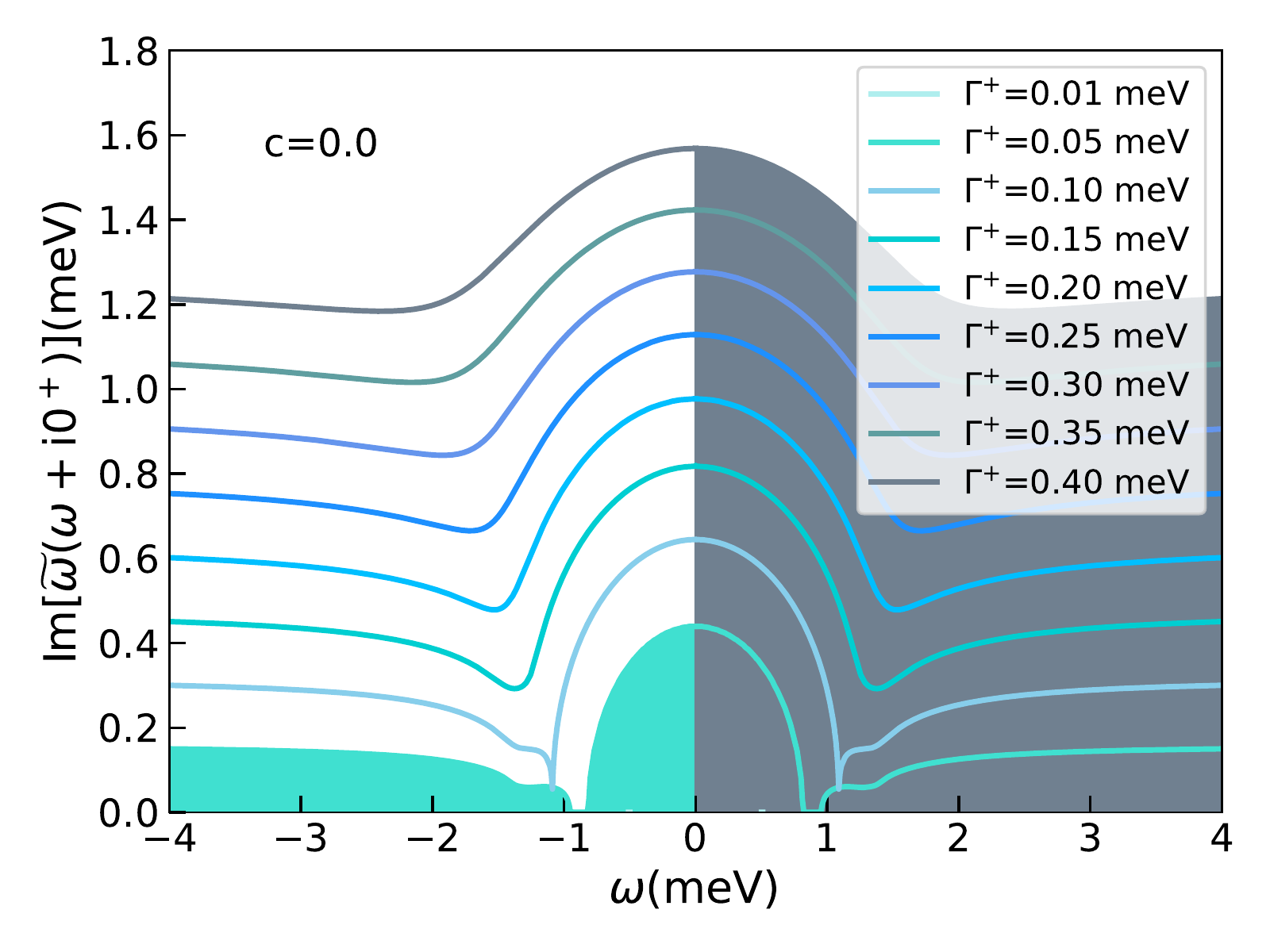}
\caption{Evolution of the family of Wigner probabilistic distributions obtained from imaginary part of the scattering cross-section in the unitary limit, for eight values of non magnetic concentration parameter $\Gamma^+$(meV) in the RPS for a quasi-point model and the tiny MN gap.}
\end{figure}

\section{The unitary limit for the case of nodal points and a flat dispersion law in the FS gamma-sheet of strontium ruthenate, and the metallic half filling ground TB state.}

If $|\epsilon_F|$ $\approx$ $\frac{|t|}{10}$, i. e., $|\epsilon_F|$ $\ll$ $|t|$, the unitary regime and the family of Wigner distribution functions of figure 4, show a different behavior from the one studied in the previous section. It resembles a gapless node points in unconventional superconductors (also it resembles the line nodes points model of high Tc compounds we previously calculated for $La_{2-x}Sr_xCuO_4$ in [29]). This intuitively means that the half filling ground state in the unitary limit gives point nodes and a flat dispersion law if the TB parameters are appropriately chosen.

Therefore, the normal state quasiparticles with a flat $\gamma$-sheet in the unitary region for $Sr_2RuO_4$, and with a gapless behavior have also an ill-defined momentum quasiparticles between elastic collisions, can be studied using the scattering cross-section following the same methodology. The signature of the unitary state still corresponds to the resonance at zero frequency in the imaginary part of the scattering cross-section, as it can be seen from the Wigner distribution probabilities functions, obtained self-consistently in figure 4, but in this case there is not a tiny gap as in the previous section, as should be, since the superconducting regime is gapless in the case of point nodes.

In this case, there is only one macroscopic phase in $Sr_2RuO_4$, for both, the  dilute levels of non-magnetic disorder composed by Cooper pairs and the normal state quasiparticles (turquoise line and region in phase scattering space shaded turquoise in figure 4), and where the following identity holds: $(\tilde{\omega} \big(\omega+ i 0^+) \neq 0 \, meV)$, for the whole range of energies from 0 to 4 meV (with a minimum region shaded turquoise, and maximum region in scattering reduced phase space shaded gray in figure 4. 

This is a new class of Wigner probabilistic distributions family found for $Sr_2RuO_4$ in the case of point nodes for the $\gamma$-flat Fermi sheet, and it shows how numerically depending on the values of the TB parameter $\epsilon_F$ there can be point nodes in this compound as well, and not only a tiny gap family of Wigner distribution functions [20]. We wish to point out, that the use of the symmetric point group D$_{4h}$ and the symmetry broken state in the case of strontium ruthenate is also supported by recently review works [37-39], and the flat bands approach is widely discussed for heavy fermions compounds in the monograph [40].

\begin{figure}[ht]
\includegraphics[width = 6.0 in, height= 5.0 in]{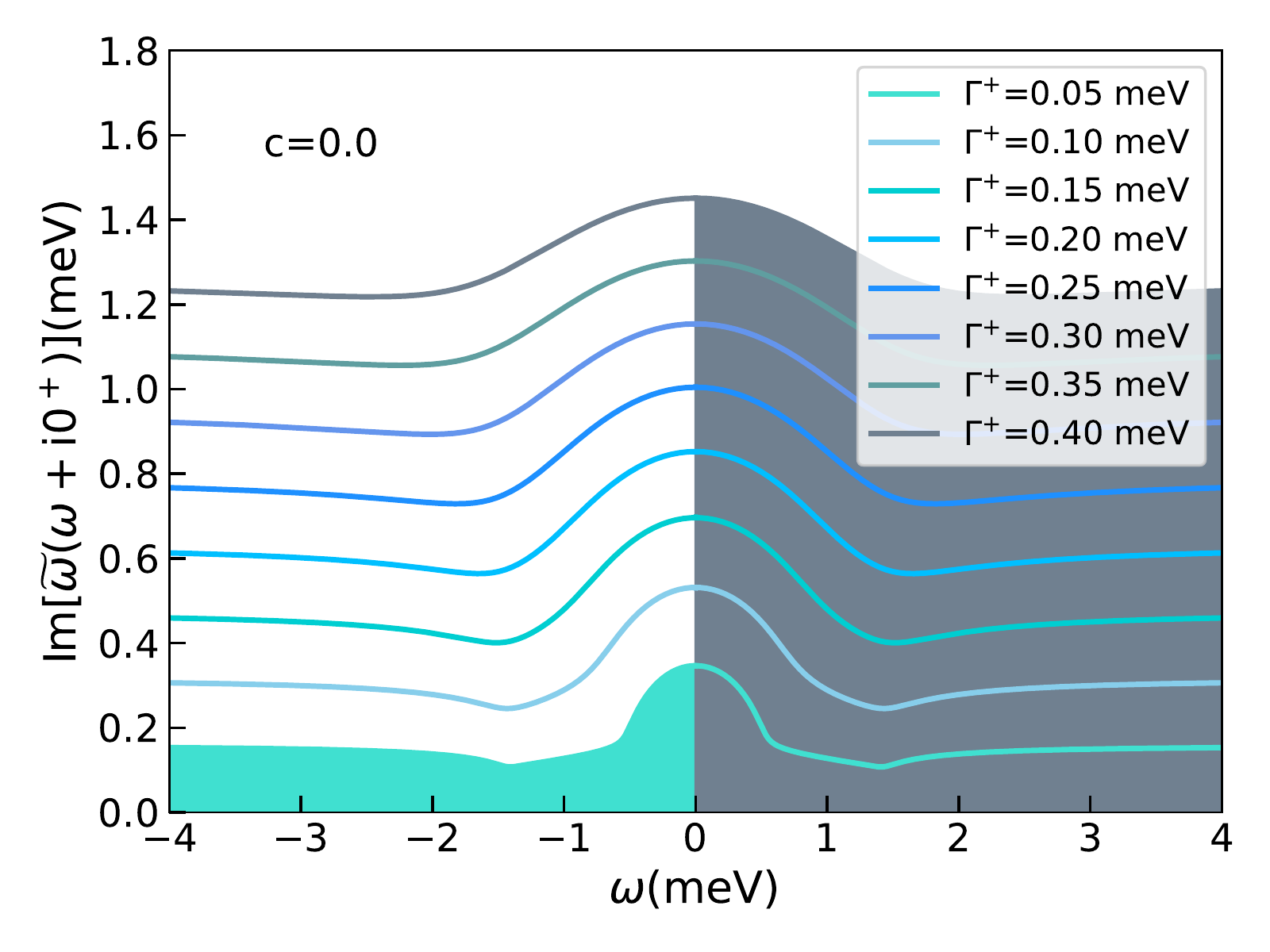}
\caption{Evolution of the family of Wigner probabilistic distributions obtained from the imaginary part of the scattering cross-section in the unitary limit, for nine values of non magnetic disorder parameter $\Gamma^+$(meV) in the scattering reduced phase space. for a point lines mode with a flat dispersion law and a metallic half filling ground state.}
\end{figure}

\clearpage

\section{Conclusions}

This communication was aimed at investigating numerically the behavior of the non-magnetic disordered imaginary part of the elastic scattering cross-section in the unitary metallic limit for two cases in the FS $\gamma$-sheet of $Sr_2RuO_4$. One case with TB parameters for the dispersion law $\xi_\gamma(k_x,k_y)$ given by $|\epsilon_F|$ $\approx$ $|t|$ with represents a dispersed quantum mechanical (QM) electronic behavior, and the case with $|\epsilon_F|$ $\ll$ $|t|$ with a flat $\gamma$ QM sheet behavior.

First we found and discuss an interesting feature, the MN tiny gap model at half filling in the $\gamma$ sheet disappears, and converts into a point nodes case in the RPS with D$_{4h}$ symmetry, if in the TB first neighbor approximation, we use for an almost flat sheet the parameter value $|\epsilon_F| = 0.04 \, meV$, otherwise is $|\epsilon_F| = 0.4 \, meV$ for quasinodal behavior. In figures 1 and 2 we illustrate and compare the two cases mentioned above, using an implicit function scheme for both, the dispersion law and gap symmetry equations with the corresponding TB parameters. In addition in section 2, we briefly introduced the Edwards-Nambu-Gorkov formalism needed for a TB analysis of the RPS. 

Second, in Figure 3, we numerically calculate and sketch a first family of eight Wigner distributions, where the region corresponding to optimal levels of disorder, i.e., $\Gamma^+$ = 0.4 meV, has been shaded gray in the right side of the plot. In addition, the regions corresponding to dilute levels of disorder with $\Gamma^+$ = 0.05 meV, has been shaded turquoise in the left side of the plot. The Miyake-Narikiyo tiny gap that corresponds to 4 quasi-nodal points in the order parameter, and graphically is showed in figure 3 as a turquoise line in the left side.

Third, in section 4, the behavior of the disordered non magnetic matrix $\Im[\tilde{\omega}\big](\,\omega+i\,0^+)$ inside the unitary ($c$ = 0) RPS was studied for eight values of $\Gamma^+$, starting at very diluted disorder, optimal disorder values, and finally an enriched disorder concentration. In Figure 4, the region corresponding to optimal levels of disorder, i.e., $\Gamma^+$ = 0.4 meV, has been shaded gray in the right side of the figure. In addition, in this case the MN tiny gap does not appear in the imaginary part of the scattering cross-section, i.e., it represents the case of the 4 point nodes with a flat FS $\gamma$-sheet, which is shaded turquoise in figure 4 on the left side, and corresponds to the turquoise line at value $\Gamma^+$ = 0.05 meV. 

We find according to figure 3, that for the case of the MN tiny gap, there are two phases: a phase in the RPS with doping $\Gamma^+$ $\geq$ 0.1 meV where Fermion dressed RPS quasiparticles are mixed with Cooper pairs. This can be seen in Figure 3 (gray region). We also found the MN tiny gap phase for $\Gamma^+$ $=$ 0.05 meV, in that region only Cooper pairs exist for an energy window between 0.85 meV and 1.0 meV as is seen in figure 3, left side (turquoise line) as noticed previously [20].

Also, we find according to figure 4, that for the case point nodes, exists macroscopically only one phase with both, Cooper pairs and normal state point nodal quasiparticles, reflecting a new family of Wigner probabilistic distributions in the RPS with doping $\Gamma^{+}$ $\geq$ 0.05 meV, for the whole energy window, i.e., (-4 meV, +4 meV), this can be seen in Figure 4 (turquoise and gray regions). This phase resembles High Tc cuprates. In the case, we conclude that there is always a minimum amount of Fermi dressed quasiparticles interacting elastically due to a small amount of non-magnetic disorder $\Gamma^+ = 0.05$ meV, the main difference with lines nodes case [29] is that here, $\Delta_0^{\gamma}$ $= 1.0$ meV (an smaller RPS window (-4.0 to +4.0 \; meV [20]) and in the line nodes high T$_c$ case, $\Delta_0^{\gamma}$ $= 33.9$ meV (bigger RPS window (-33.9 to +33.9 \; meV [20,29]) .

We end this communication pointing out that the Miyake-Narikiyo gap expression using a tight binding model is very useful for setting up numerical studies in triplet superconductors such as $Sr_2RuO_4$, as we have demonstrated in this and previous studies on the subject [19-20]. This approach uses a macroscopic Wigner distribution probabilities phenomenological approach obtained from the analysis of the imaginary part of the scattering cross-section, helping undoubtedly to resolve the actual conjecture of the location of the nodes in $Sr_2RuO_4$.

Our studies involve using several families of TB Wigner distribution probabilities with the Miyake Narikiyo model obtained self-consistently, aiming at clarifying the location of point nodes in the FS $\gamma$-sheet broken symmetry state. They could be used to study other irreducible representations belonging to different point symmetry groups in other crystalline structures, so far, we have considered 2 of them, remaining several more to be studied. 

The existence of a link of our phenomenological TB approach with the quasi-classical approach to the strong coupling theory in unconventional superconductors proposed in [42] and recently discussed in [43] could improve the development of the use of Wigner distributions probabilities [18,44] in order to study new unconventional superconductors with a reduced scattering phase space with dressed quasiparticles and their quantum DOS states. 

\newpage

\section*{7. Acknowledgements}
We thank conversations with Dr E. Beliayev and Prof. Sh. M. Ramazanov about the use of quantum distribution functions to study
macroscopic properties in unconventional superconductors. Also we thank an anonymous reviewer whose technical comments helped improve and clarify this manuscript. The authors did not receive any financial support for research, authorship and/or publication of this article.

\end{document}